\documentclass[aps,prd,reprint,preprintnumbers,superscriptaddress,nofootinbib,longbibliography]{revtex4-2}
\usepackage[utf8]{inputenc}
\usepackage{amsmath,amssymb,array,calc,rotating,epsfig,psfrag}
\usepackage{braket}
\usepackage{natbib,hyperref}
\usepackage[capitalise]{cleveref}
\usepackage{float}
\usepackage{tensor}
\usepackage{svg}
\usepackage{xspace}
\usepackage{lineno}
\usepackage{graphicx}
\usepackage{subcaption}

\begin{document}
\preprint{FERMILAB-PUB-26-0553-CSAID}
\title{Confinement, String Breaking, and Hadronization in the Compact Abelian Higgs Model}

\author{Blake Senseman}
\affiliation{The University of Iowa, Iowa City, IA 52242, USA}
\author{Zane Ozzello}
\affiliation{The University of Iowa, Iowa City, IA 52242, USA}
\author{Yannick Meurice}
\affiliation{The University of Iowa, Iowa City, IA 52242, USA}
\author{Stephen Mrenna}
\affiliation{Fermi National Accelerator Laboratory, Batavia, IL 60510, USA}

\date{\today}

\begin{abstract}
While real-time simulation of Quantum Chromodynamics remains technologically out of reach, many simplified models for studying elements of QCD phenomenology are being investigated. In this work, we present a simple model, a spin-1 truncation of the Compact Abelian Higgs Model simulated on qutrit sites, in which confinement and string breaking are accessible to current simulation methods. In the low-energy regime of 1+1 dimensional scalar electrodynamics, the heavy modes are integrated out, producing a spin chain effective Hamiltonian in which Gauss' law is automatically satisfied. We study the spectrum of string-like excitations using DMRG methods on the order of 100 sites. We demonstrate that an added, local chemical potential, playing a role analogous to external charges, permits parameter-dependent measurements of physical features of interest like the string tension and effective meson mass. Varying the chemical potential also permits a characterization of string stability not assessed in prior studies of confining lattice models.
\end{abstract}

\maketitle

\section{Introduction}

The emergence of hadrons from quarks and gluons in high-energy collisions remains one of the most challenging phenomena to describe quantitatively within quantum chromodynamics (QCD). While the short-distance dynamics of partons can be treated perturbatively, hadronization occurs in the strongly coupled regime where analytic methods are limited. Phenomenological models such as the Lund string framework \cite{Andersson1983StringFragmentation} implemented in event generators like \textsc{Pythia} \cite{sjostrandPYTHIAPhysicsManual2006a, Bierlich2022Pythia83} have achieved remarkable success in reproducing experimental data, but they rely on classical effective descriptions where parameters governing the shape of fragmentation functions, relative frequency of produced species, and beam remnant properties (among others) are determined largely by statistical fitting. 

Lattice gauge theories, on the other hand, offer a natural framework in which confinement and flux-tube formation arise dynamically from microscopic degrees of freedom \cite{BanulsCichy2020Review}. Recent advances in tensor network methods and programmable quantum simulators have made it possible to study their real-time evolution \cite{Kuhn2015NonAbelian, Farrell2024HadronDynamics}. Recent, leading studies have used a variety of gauge groups, Hamiltonian representations, and computational methods to identify string phenomenology in 1+1 dimensional \cite{De2024, Liu2025, Ciavarella2024, Mildenberger2025Z2Confinement, Alexandrou2025Z2StringBreaking} and 2+1 dimensional \cite{Cochran2025, GonzalezCuadra2025, Gupta2026StringBreakingSU2, Cobos2025krn, Xu2025abo, Borla:2025gfs, Cataldi2025RealTimeString} lattice systems. Many energetic and entropic measures have been used to identify the proliferation of confined quasiparticles \cite{Senseman:2026Prolegomena, Vovrosh2021, calabreseEvolutionEntanglementEntropy2005, kormosRealtimeConfinementFollowing2017}, but perhaps the most consistent characterization between different models is the effective potential induced between external charges \cite{Liu2025,GonzalezCuadra2025,Gupta2026StringBreakingSU2,Surace2026StringBreakingDynamics}. These developments motivate the investigation of string formation and fragmentation in lattice models as a potential bridge between first-principles gauge theory dynamics and the statistical features of hadron production observed in high-energy experiments \cite{Meurice2023ScalarQED}.

This work aims to demonstrate that the spin-1 truncation of the Compact Abelian Higgs Model (CAHM) hosts string-meson modes whose spectrum can be calculated in appropriate limits, and that physical parameters of interest (effective string tension, effective meson masses) can be calculated by identifying minimal-energy states using density matrix renormalization group (DMRG) methods \cite{Schollwock2011DMRG} in the presence of an additional chemical potential. Section II will define the Hamiltonian for the CAHM, and discuss its spectrum of string-meson excitations. Section III will present the results of the DMRG study of the model demonstrating (1) a universal linear potential for external charges, (2) measurable string tension and effective meson mass, and (3) stability of these results under variation of the chemical potential.

\section{The Truncated Compact Abelian Higgs Model}
The target model of this work is a truncation of the Compact Abelian Higgs Model (CAHM), which is a low-energy effective description of 1+1 dimensional scalar electrodynamics \cite{Meurice2021AbelianHiggs}. This model has previously been implemented on a chain of spin-1 sites using tensor network methods \cite{Zhang2018PolyakovLoop}, on a $2 \times N$ array of Rydberg atoms \cite{Senseman:2026Prolegomena}, and on a small number of transmon qutrits \cite{Asaduzzaman2026HybridAnalogDigital}. 
The Hamiltonian for a 1D chain CAHM is
\begin{equation}
\label{eq:CAHM_ham}
     H_{CAHM} = \frac{U}{2}\sum_{i} (L^z_{i})^2 
+ \frac{Y}{2} {\sum_{<ij>}}   (L^z_{i} - L^z_{j})^2-
\frac{X}{\sqrt{2}} \sum_{i} L^x_{i}
\end{equation}

\noindent where $L^z$ and  $L^x \equiv \frac{1}{2}(L^+ + L^-)$ are spin operators acting on the site-wise quantum number $z$, which is truncated to take on values on $(-m_{max}, m_{max})$.
\begin{align}
    L^z \ket{m} &= m \ket{m}, \quad L^\pm \ket{m} = \ket{m\pm 1}, \\\nonumber
    & \quad L^\pm \ket{\pm m_{max}} = 0
\end{align}

For this work, $m_{max}=1$ which gives a formulation on spin-1 (qutrit) sites. Such a tight truncation is used, not as an approximation to the continuum Abelian-Higgs model, but because it is very amenable to interpretation along the lines of Lund string dynamics. 

The $m$ quantum number plays the role of an electric field which, when excited to $m=\pm 1$, carries an energy per site of $U/2$. This means that pairs of domain walls (changes in the value of $m$ along the chain) experience a linear potential, manifesting confinement as studied in many quantum lattice models \cite{De2024, Liu2025, Ciavarella2024, Cochran2025, GonzalezCuadra2025}. The second term in the Hamiltonian grants an energy of $Y/2$ to each domain wall, which can be understood as a combined mass-coupling parameter for the implicit charged matter, the distribution of which can be inferred by applying Gauss' law to the electric field representation. The presence of such an energy cost for creating domain walls controls the tendency for string breaking. Both string motion and string breaking are achieved by the $X$ term, which is the only operator not diagonal in the electric field basis. We conventionally set $X=1$, which determines the numerical energy scale for the Hamiltonian.

\subsection{Confined Quasiparticles (High Y Limit)}
\label{sec:quasiparticles}
The confinement of domain walls caused by the constant energy density of the electric field produces fluctuating, string-like quasiparticles. Their spectrum is most naturally accessed in the limit $Y \xrightarrow{} \infty$. There the model spectrum breaks up into subspaces characterized by the number of domain walls. If we additionally require that the field vanish on the boundary ($m_i=0 \text{ for } i = 1,N$) or use periodic boundary conditions, then the two-domain-wall subspace is recognizable as two separated copies of the Hilbert space of a 1D lattice string, one for $m=1$ and an identical one for $m=-1$. These string subspaces are swapped by charge conjugation, under which the model Hamiltonian is invariant.
\begin{align}
    C L^z C^{-1} &= -L^z, \quad C L^x C^{-1} = L^x \\\nonumber
    & \Rightarrow C \; H_{CAHM}  \; C^{-1} = H_{CAHM}
\end{align}
If the model Hamiltonian is restricted to one of these string subspaces with the appropriate projector $P_s$, the effective string Hamiltonian contains one term that attributes an energy linear in the length of the string and another term that accomplishes hopping of the location of the endpoints, labeled with lattice site indices $(i,j)$ with $i<j$.

\begin{align}
\label{eq:string_Ham}
\hat{H}_{s}
\equiv & P_s H_{CAHM} P_s \\
= &Y \; + \; \sum_{i<j}^L\frac{U}{2} (j-i)\ket{i,j}\bra{i,j} \nonumber\\
&- \frac{X}{2} \bigg( \ket{i+1,j}\bra{i,j} + \ket{i,j+1}\bra{i,j} + \mathrm{h.c.} \bigg)\nonumber
\end{align}

This Hamiltonian has been diagonalized \cite{suraceScatteringMesons2021} to produce a basis of string-meson momentum eigenstates using quantum numbers $(k,\ell)$ that arise from the momentum conjugate to the center of mass coordinate $s\equiv i+j$ and the boundary conditions on the relative coordinate $r\equiv j-i$, respectively. For the spin-1 truncated CAHM, those wavefunctions contain Fourier and Bessel ($\mathcal{J}_n(q)$) modes with the following form \cite{suraceScatteringMesons2021}.
\begin{equation}
    \ket{\Psi_{k,\ell}} = \sum_{i<j}^L e^{iks}\mathcal{J}_{r-\nu_{k,\ell}}(4X \cos{k}/U) \ket{i,j}
\end{equation}\begin{equation}
    \nu_{k,\ell} \equiv \ell\text{-th root of } x \xrightarrow{} \mathcal{J}_{-x}(4X \cos{k}/U)
\end{equation}
These wavefunctions span the one-string subspace and will be mixed and dressed with multi-particle corrections when a finite $Y$ is used. In the rest of this paper, we use a finite $Y$ to facilitate string breaking.

\subsection{Chemical Potential for String/Meson States}

\begin{figure}[h!]
    \centering
    \includegraphics[width=1.1\linewidth]{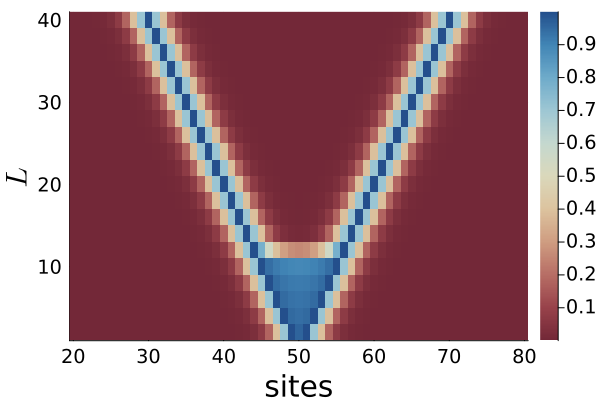}
    \caption{Expected field value $\langle L^z \rangle$ for the ground state with $U=0.2, Y=2.0$ and a local chemical potential $(\mu=100)$ applied at sites separated at different lengths $L$. The ground state displays a string configuration up to $L=10$, after which point a broken string state containing two, independent localized mesons is preferred.}
    \label{fig:breaking}
\end{figure}

\begin{figure*}[t!]
    \centering
    \includegraphics[width=0.49\textwidth]{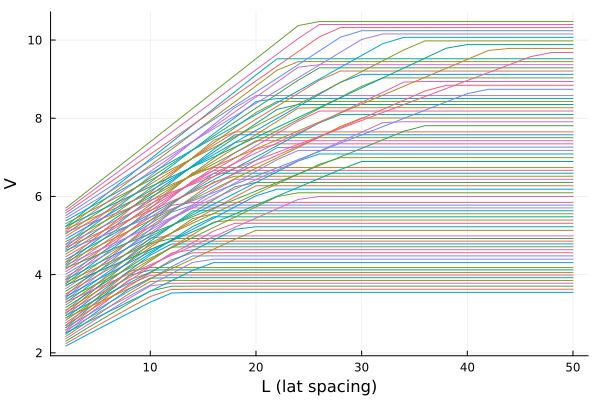}
    \hfill
    \includegraphics[width=0.49\textwidth]{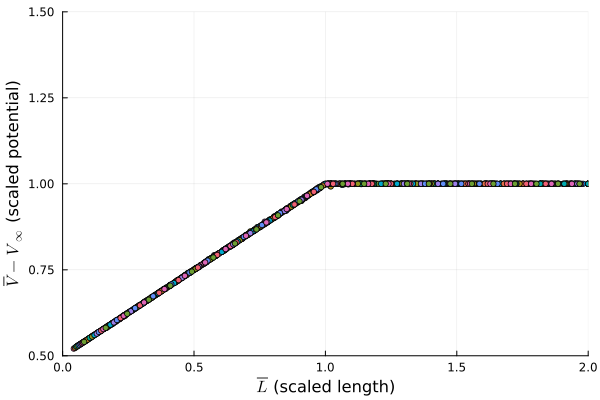}
    \caption{Measured string potentials $V(L; U,Y)$ (left) for $Y \in (2.0, 2.5, ..., 6.0), U \in (0.1, 0.125, ..., 0.35)$ obtained with $\mu=100$. The ground state energy $E_0(U,Y)$, calculated with $\mu=0$, has been subtracted from each sample. Rescaling potentials (right) demonstrates the universal nature of the string behavior once $\sigma$ and $L^*$ are observed. A suitable vertical shift has also been performed to account for the discrepancy between $M_{eff}$ and $\sigma L^*$.}
    \label{fig:potentials}
\end{figure*}

In order to probe the spectrum of string-like excitations, it has become conventional to insert external charges and minimize the energy of the electric field configuration subject to that condition \cite{Liu2025,GonzalezCuadra2025,Gupta2026StringBreakingSU2,Surace2026StringBreakingDynamics}. However, since this model most naturally exposes electric field degrees of freedom, this study adds a strong, local chemical potential term to a set of sites $\{s\}$ in order to drive the field into an excited state. \footnote{This perturbation also has the form of an interaction with a fixed dipole $-$\textbf{E} $\cdot$ \textbf{p}, but we use the terminology ``local chemical potential" for simplicity.}
\begin{equation}
    H' = H_{CAHM} - \mu \; L^z_{\{s\}}
\end{equation}

This is analogous to the insertion of external charges adjacent to those sites, which would also produce $m=1$.
However, our perturbation of the Hamiltonian introduces a parameter $\mu$ that can be used to study the impact of varying the perturbation, which is not available when external charges are inserted.

\section{Static String Analysis}

To study the string potential, energy-minimizing states are obtained by DMRG with the local chemical potential applied to two sites with varying distance $L$ between them. Such states generally have the appearance of nearly constant electric flux between the two endpoints up to a certain length $L^*$, after which the energetically-preferred state contains localized excitations around the sites where the local chemical potential is applied. These will be referred to as ``string" and ``broken" states, respectively. The transition from string to broken states as $L$ changes can be seen in Figure \ref{fig:breaking}. All DMRG work is done with 40 sweeps and a maximum bond dimension of 200.

\subsection{String Potential, Tension, and Breaking Length}
\label{sec:params}

The string potential $V(L; U,Y, \mu)$ is defined as the excess in energy of the length-dependent states above the unperturbed model ground state. Plotting the potential over a wide range of model parameters in Figure \ref{fig:potentials} reveals the saturating linear potential observed in prior studies \cite{Bergner2026AdjointQCD,Crippa2026ConfinementString,Buyens2016Confinement,Surace2026StringBreakingDynamics}. Each sample is fit with a continuous, piecewise linear function, yielding the string tension $\sigma$ as the slope of the first segment and the breaking length $L^*$ as the position of the knee. The effective string tension exceeds the microscopic value $(U/2)$ because the string is dressed with virtual domain wall pairs; the resulting matter density along the string is discussed in Section \ref{sec:varying_mu}.

Rescaling the potentials and the lengths using the relations below shows the striking universality of this linear potential model over the chosen parameter space.
\begin{equation}
    \overline{V} \equiv V / (2\sigma L^*), \quad \quad \overline{L} \equiv L / L^*
\end{equation}
The string tension and breaking length can be seen in Appendix \ref{app:params} to vary smoothly over the parameter range of this study.

\subsection{Localized Meson Mass Scale}
\label{sec:energy scales}

\begin{figure}[t!]
    \centering
    \includegraphics[width=\linewidth]{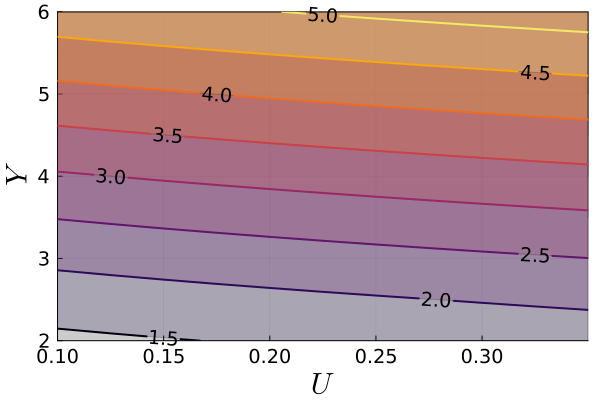}
    \caption{Effective meson mass scale $M_{eff}$ for $Y \in (2.0, 2.5, ..., 6.0)$, $U \in (0.1, 0.125, ..., 0.35)$ calculated with DMRG}
    \label{fig:M_eff}
\end{figure}

The introduction of the local chemical potential on a single site, when subject to energy minimization through DMRG, yields localized excitations, the energy of which is related to the mass scale of meson excitations. Indeed, within the studied parameter range, the ratio of the localized excitation energy $M_{eff}$ to the finite-volume mass gap $\Delta_m$ for string-like excitations in the model is very nearly unity, with a leading-order inverse power law correction in $Y$. The localized energy scale is shown in Figure \ref{fig:M_eff} to be remarkably linear in $Y + U/2$, which would be the coefficient of $(L^z)^2$ if the $Y$-term in Eq. (\ref{eq:CAHM_ham}) were expanded.

The classical picture of string breaking suggests that a string will break when the energy added to the minimum energy meson is sufficient for the creation of a second meson, or in other words, that
\begin{equation}
    M_{eff} + \sigma L^* = 2M_{eff}.
\end{equation}

\begin{figure}[h!]
    \centering
    \includegraphics[width=\linewidth]{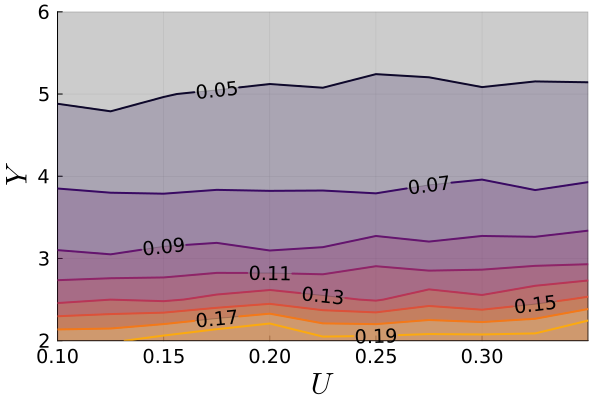}
    \caption{Error associated with deviation from simple string breaking model calculated as  $M_{error} = M_{eff} - \sigma L^*$. The percent difference goes from $12\%$ at $Y=2.0$ to $<1\%$ at $Y=6$, and is nearly independent of $U$.}
    \label{fig:M_error}
\end{figure}

In quantum lattice models, this coincidence in the spectrum is associated with the phenomenon of resonant string breaking \cite{GonzalezCuadra2025, Liu2025, Verdel2020StringBreaking, Cataldi2025RealTimeString, Xu2025abo}, where transitions between these states are strongly enhanced. The applicability of semiclassical string dynamics to the model can therefore be assessed by its adherence with this condition. The deviation, plotted in Figure \ref{fig:M_error} as
\begin{equation}
    M_{error} \equiv M_{eff} - \sigma L^* ,
\end{equation}
drops off rapidly with string stiffness ($\sim Y^{-1.45}$) and is very nearly independent of $U$. Further comparisons of the parametric behavior of $\Delta_m, M_{eff}, \text{ and } \sigma L^*$ are available in Appendix \ref{app:masses}. These observations further justify the supposition from the effective string spectrum that higher $Y$ yields more recognizable and consistent semiclassical string behavior. 

\subsection{Adiabatic Removal of the Chemical Potential}
\label{sec:varying_mu}

The strong, local chemical potential used to construct string states in this model, which is analogous to the inclusion of external charges next to those sites, has shortcomings from the perspective of preparing physically plausible string states. By forcing the field at that site to take on its maximum value, it interrupts the natural transition between the bulk and edge of the string. Indeed, one might wonder whether the string state is natural within the unperturbed spectrum since the added term is so significant.

These concerns can be addressed by adiabatically decreasing $\mu$ through successive rounds of DMRG. Since the broken state is non-locally distinct from the string state, it is quite distant in the space of matrix product states, and the DMRG procedure can be guided through a local minimum maintaining the extent of the string state. This generally makes a mild change to the energy of the string state, and so can introduce corrections to the inferred string parameters $(\sigma, L^*)$. 

\begin{figure}[h!]
    \centering
    \includegraphics[width=\linewidth]{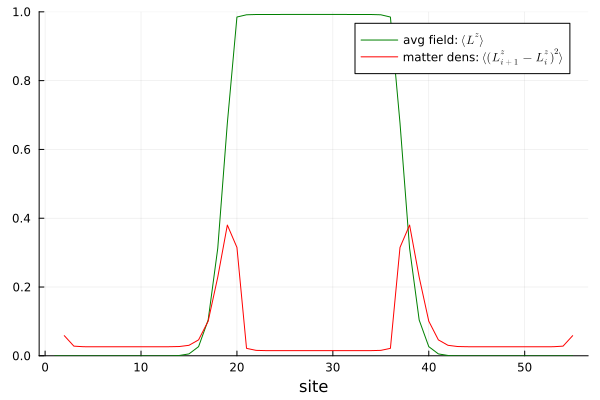}
    \caption{The spatial profile of the average field and matter density for an adiabatically prepared string state ($U=0.25$, $Y=6.0$, $L=16$, $\mu=2.25$). Both the string interior and the vacuum exterior can be seen to be dressed with virtual domain wall pairs.}
    \label{fig:string_profile}
\end{figure}

\begin{table}[ht]
\centering
\resizebox{0.65\columnwidth}{!}{
\begin{tabular}{c|ccc}
$\mu$ & $x_i=20$ & $x_i=21$ & $x_i=22$ \\
\hline
100.0 & 1.0000 & 0.9927 & 0.9922 \\
10.0  & 0.9985 & 0.9926 & 0.9922 \\
5.0   & 0.9955 & 0.9923 & 0.9922 \\
3.0   & 0.9897 & 0.9920 & 0.9922 \\
2.5   & 0.9859 & 0.9915 & 0.9922 \\
2.25  & 0.9830 & 0.9909 & 0.9922 \\
\end{tabular}}
\caption{Measured $\langle L^z \rangle$ at three different sites $x_i=(20,21,22)$ throughout the process of adiabatic preparation with decreasing $\mu$.}
\label{table:mu}
\end{table}

\newpage
To illustrate this, Figure \ref{fig:string_profile} depicts the average field and matter density of a string with $U=0.25$, $Y=6.0$, $L=16$ where the local chemical potential has been removed slowly from $\mu=100$ to $\mu=2.25$, which is the minimum value at which we were still able to preserve the string state. This adiabatic procedure decreased the vacuum-subtracted energy of the string state by 0.1 (in $X$ units), or by about $1.5\%$. This change is negligible for a cursory study of the properties of the model, but could become relevant for precision study of the avoided crossing associated with string breaking \cite{Surace2026StringBreakingDynamics}.
Table \ref{table:mu} shows how the declining $\mu$ removes the artificially elevated edge of the string for the $L=16$ string in Figure \ref{fig:string_profile}.

\section{Conclusion}
Both exact solutions of the model Hamiltonian in the appropriate limit and recovery of a consistent potential model in DMRG demonstrate the existence of string-meson phenomenology in the spin-1 truncation of the Compact Abelian Higgs Model. Physically relevant parameters like the string tension and effective meson mass can be inferred from DMRG with local chemical potential terms simulating the inclusion of external charges. These external charges can be adiabatically weakened well below the energy scale of the strings themselves without losing the structure of the string state, leading to more precise representation of localized, stationary strings within the model. These methods may also be beneficial in the creation of physically plausible string states that can be used as initial states or as part of a measurement procedure for future analysis of the time-evolution of strings, including spontaneous string breaking. We hope that future work will be able to observe dynamical string breaking behavior using real-time evolution methods in this and other models. Such work may shed light on elements of phenomenological models used by event generators prior to fitting with collider data.

\begin{acknowledgements}
B.S. and Z.O. acknowledge support from the URA Visiting Scholars Program. 
B.S., Z.O., and Y.M. are supported in part by the Dept. of Energy under Award Numbers DE-SC0019139, DE-SC0026494, and DE-SC0010113. 
This work benefited from the workshop
``From String Dynamics to Event Generation with Quantum Simulation" (April 2026), 
which was supported by U.S. Department of Energy, Office of Science, Office of Nuclear Physics, InQubator for Quantum Simulation (IQuS) under Award Number DE-SC0020970 via the program on Quantum Horizons: QIS Research and Innovation for Nuclear Science.
This manuscript has been authored by Fermi Forward Discovery Group, LLC under Contract No. 89243024CSC000002 with the U.S. Department of Energy, Office of Science, Office of High Energy Physics.


\end{acknowledgements}

\newpage

\bibliographystyle{apsrev4-1}
\bibliography{dmrg_bib.bib}

\newpage

\newpage

\appendix
\section{String Model Calculations in DMRG}
\label{app:params}

\begin{figure}[h!]
    \centering
    \includegraphics[width=\linewidth]{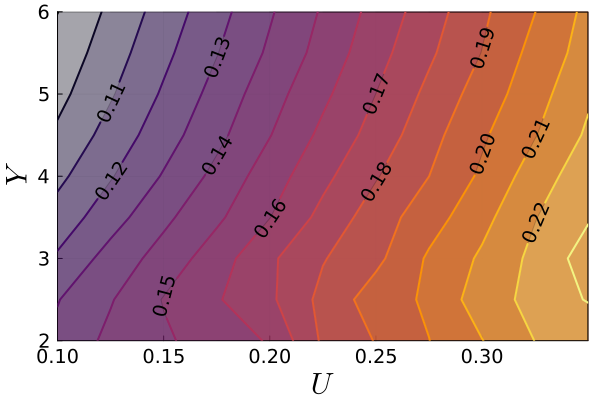}
    \caption{String tension $\sigma$ for $Y \in (2.0, 2.5, ..., 6.0)$, $U \in (0.1, 0.125, ..., 0.35)$ calculated from the slope of the potential}
    \label{fig:str_tens}
\end{figure}

\begin{figure}[h!]
    \centering
    \includegraphics[width=\linewidth]{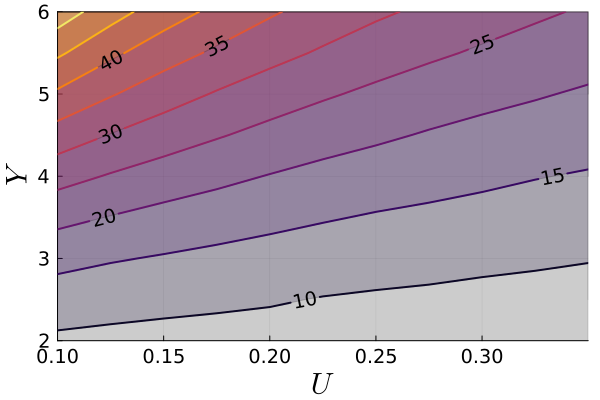}
    \caption{Breaking length $L^*$ for $Y \in (2.0, 2.5, ..., 6.0)$, $U \in (0.1, 0.125, ..., 0.35)$ calculated from the location of the knee in the potential}
    \label{fig:break_pts}
\end{figure}

The effective string tension and the string breaking length can be extracted from the very general form of the static string potential defined in Section \ref{sec:params} by a piecewise linear fit. These parameters show a remarkably smooth dependence on the Hamiltonian parameters in the range studied. The string tension can be seen in Figure \ref{fig:str_tens} to be linear in $U$ for high Y, as would be expected from the microscopic dynamics. The breaking length, as expected, is largest with heavy domain walls (high $Y$) and weak field (low $U$). The range of lengths in Figure \ref{fig:break_pts} demonstrate the reason for the chosen parameter range -- strings shorter than the localized meson excitations ($\sim 10$ sites) fail to display a linear potential, and much larger strings would be too high a fraction of the simulation volume.


\section{Comparison of mass parameters}
\label{app:masses}

In Section \ref{sec:energy scales} above, three measures of the effective mass of string-meson quasiparticles are defined. The truest is the finite-volume meson mass gap $\Delta_m$, which is calculated in DMRG with a projective exclusion of the previously computed ground state and an added energy penalty for boundary excitations. This produces the lowest energy state within the one-string sector. The localized meson energy $M_{eff}$ is calculated as the vacuum-substracted DMRG minimum for a local chemical potential applied to a single site. The behavior of the potential also yields a mass scale governing string breaking calculated as $M_\sigma \equiv \sigma L^*$.

All three measures show remarkable similar behavior as a function of model parameters. One convenient fit is to a linear model
\begin{equation}
    M = a \; (Y+U/2) + b \; (Y-U/2) + c
\label{eq:linear}
\end{equation}

which makes apparent the dominant dependence on the $(L^z)^2$ term in the model Hamiltonian if the $Y$-term in equation \ref{eq:CAHM_ham} were expanded. 

\begin{table}[h!]
\resizebox{0.7\columnwidth}{!}{
\begin{tabular}{|l|l|l|l|l|}
\hline
\textbf{}                                  & \textbf{a} & \textbf{b} & \textbf{c} & \textbf{$1-R^2$} \\ \hline
$\Delta_m$                   & 1.608      & -0.175     & -0.964     & 2.1e-4                          \\ \hline
$M_{eff}$                 & 1.681      & -0.062     & -0.629     & 2.2e-4                          \\ \hline
$M_\sigma$ & 1.70       & -0.110     & -0.845     & 9.4e-4                          \\ \hline
\end{tabular}
}
\caption{Linear fit of meson mass scales using the model in equation \ref{eq:linear}}
\end{table}

\begin{figure}[h!]
    \centering
    \includegraphics[width=1.1\linewidth]{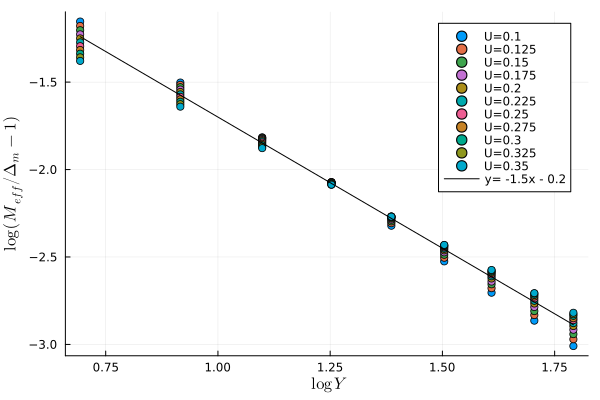}
    \caption{Comparison of $M_{eff}$ and $\Delta_m$ as a function of $Y$ for each value of $U$, and an averaged linear fit (black line). The leading order power-law tail appears to have a mild $U$-dependence.}
    \label{fig:M_eff_comp}
\end{figure}

\begin{figure}[h!]
    \centering
    \includegraphics[width=1.1\linewidth]{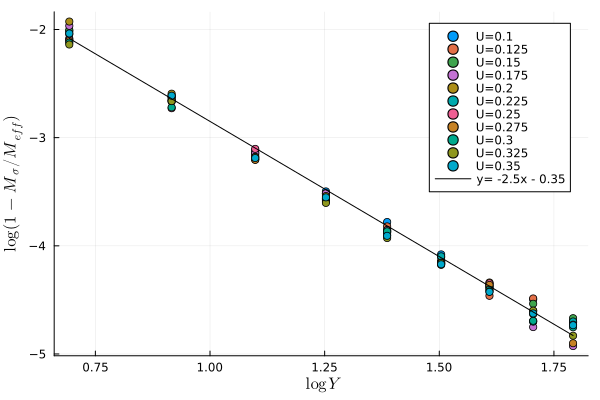}
    \caption{Comparison of $M_\sigma$ and $M_{eff}$ as a function of $Y$ for each value of $U$, and an averaged linear fit (black line). The leading order power-law tail appears to have a mild $U$-dependence.}
    \label{fig:M_sigma_comp}
\end{figure}

Both $M_{eff}$ and $M_\sigma$ can be compared with $\Delta_m$ in ways that show clear asymptotic similarity in the high-$Y$ limit. The ratios $M_{eff} / \Delta_m$ and $M_{\sigma} / M_{eff}$ contain leading order corrections with the form of an inverse power law.
\begin{equation}
    M_{eff} / \Delta_m \approx 1 + 0.8 \; Y^{-1.5}
\end{equation}
\begin{equation}
    M_{\sigma} / M_{eff} \approx 1 - 0.7 \; Y^{-2.5}
\end{equation}
The parameters of both fits show some dependence on $U$, but all values of $U$ used in this study show corrections of the same form, as can be seen in Figures \ref{fig:M_eff_comp} and \ref{fig:M_sigma_comp}. These parameters are likely related to renormalization of the effective string spectrum defined in Section \ref{sec:quasiparticles}, which may be the subject of future work, but are presented here only as a limited numerical observation.

\end{document}